# Self- texturizing electronic-properties in a 2-dimensional GdAu$_2$ layer on Au(111): the role of out-of-plane atomic displacement

Alexander Correa,[a, b, c] Matteo Farnesi Camellone[d], Ana Barragan[a,c], Abhishek Kumar[e,f], Cinzia Cepek[e], Maddalena Pedio [e], Stefano Fabris [d], Lucia Vitali [a,c,g]

[a] Departamento de física de materiales, Universidad del País Vasco, ES-20018 San Sebastián (Spain)
[b] Donostia International Physics Center, ES-20018 San Sebastián (Spain)
[c] Centro de Fisica de Materiales (CSIC-UPV/EHU) y Material Physics Center, ES- 20018 San Sebastián (Spain)
[d] CNR-IOM DEMOCRITOS, Istituto Officina dei Materiali, Consiglio Nazionale delle Ricerche and SISSA, Via Bonomea 265, I-34136, Trieste, Italy
[e] Istituto Officina Materiali (CNR-IOM), Laboratorio TASC, I-34149 Trieste (Italy)
[f] Dipartimento di Fisica, Università di Trieste, I-34127 Trieste (Italy)
[g] Ikerbasque Foundation for Science, ES-48013 Bilbao (Spain)

**Here, we show that the electronic properties of a surface-supported 2-dimensional (2D) layer structure can be self-texturized at the nanoscale. The local electronic properties are determined by structural relaxation processes through variable adsorption stacking configurations. We demonstrate that the spatially modulated layer-buckling, which arises from the lattice mismatch and the layer/substrate coupling at the GdAu$_2$/Au(111) interface, is sufficient to locally open an energy gap of ~0.5eV at the Fermi level in an otherwise metallic layer. Additionally, this out-of-plane displacement of the Gd atoms patterns the character of the hybridized Gd-d states and shifts the center of mass of the Gd 4f multiplet proportionally to the lattice distortion. These findings demonstrate the close correlation between the electronic properties of the 2D-layer and its planarity. We demonstrate that the resulting template shows different chemical reactivity which may find important applications.**

Tuning the electronic properties of two-dimensional (2D) layers is a current focus of interdisciplinary investigation fields dealing with fundamental aspects of science at nanoscale and aiming to promote nanotechnological applications[1-2]. At present, one of the most promising expectations arises from the departure from the perfect crystallographic flatness of the 2D layer, typical of graphene. The emergence of new physical and chemical effects is expected from the out-of-plane displacement of the atoms in the layer and its modified orbital hybridization[3-10]. Atomic-buckling is, therefore, the fundamental parameter to tailor the electronic layer properties. In this contest, an elegant solution to template the electronic properties at nanoscale exploiting periodic atomic-buckling, is the formation of interfaces with variable atomic stacking registry, as those observed in lattice mismatched layers leading to Moiré superstructures. Notwithstanding that these Moiré superstructures are quite commonly observed at the interface of 2D-materials, little attention has been paid on the buckling-induced texturization of their electronic properties. In this work, we address the role of the out-of-plane displacement characterizing the electronic properties of a lattice-mismatched 2D-layer. We consider a 2D layer forming a Moiré superstructure and demonstrate the spontaneous patterning of the electronic properties upon the formation of the interface. We will rationalize in terms of the layer buckling and structural relaxations, the observed spatially-periodic energy shifts of the electronic structures. We will show that this variable interface coupling paves an ideal and viable route to engineer the characteristics of the 2D-layer at nanoscale leading to space modulated electronic properties and to a chemical-reaction template.

Specifically, we characterize a bi-metallic monolayer alloy, namely $GdAu_2$, obtained upon evaporation of gadolinium atoms on the annealed Au(111) surface[11-12] (see also Sup. Info.S1), and resulting in a weakly interacting Moiré superstructure. Previous works have focused on the occupied band structure of the alloy layer, characterized by photoemission experiments and Density Functional Theory calculations (DFT)[12]. In that work, the $GdAu_2$ monolayer on Au(111) was approximated as a coherent interface, neglecting the lattice-mismatch with the supporting substrate and its spatially-dependent structural relaxations. Recently, we have shown that in a similar structure, the $GdAg_2$/Ag(111), this assumption underestimates the role of the stacking geometry and that interesting electronic and magnetic effects emerge when the interface coupling is take into account[13].

Here, we combine spectroscopic techniques as scanning tunneling spectroscopy (STS), ultraviolet and inverse photoemission (UPS and IPES), with DFT calculations to achieve a precise characterization of the interface between a $GdAu_2$ layer supported on Au(111), both in the occupied and unoccupied density of states. We will show that the density of states of the alloy layer is modulated on the whole energy range with the nanoscale periodicity of the Moiré superstructure. We observe: i) the opening of a gap at the Fermi level at well-defined surface positions, ii) the change in the energy sequence of the hybridized Gd d electrons, and iii) a space-dependent energy-shift of the Gd 4f multiplet-states linearly proportional to the lattice distortions. The DFT simulations support these findings and assist the interpretation of the measured spectroscopic features.

In Figure 1, we report a topographic STM image of the $GdAu_2$/Au(111) surface showing a Moiré superlattice superposed to the atomically-resolved alloy structure. The apparent sequence of minima and maxima of the Moiré pattern reflects the variation of the vertical stacking sequence characteristics of the superposition of layers with different lattice constants. The $GdAu_2$ structure, its adsorption configurations and electronic features of the adsorbed layer have been calculated using DFT calculations. In particular, we used periodic spin-polarized DFT+U calculations that include spin-orbit coupling and a Hubbard U correction to the Gd f states (computational details in Supplementary Information). Note that the large periodicity of the superlattice prevents its direct DFT simulation because this would require exceedingly large periodic supercells. Following previous works, to keep the cost of the DFT simulations tractable, we modelled the GdAu$_2$/Au(111) system as a coherent interface [lattice parameter 5.11 Å, surface periodicity (√3 x √3)R30°] and considered three different stacking registries of the GdAu$_2$ layer with respect to the underlying Au surface. The latter was modelled with five atomic layers. These models provide realistic representations of specific Moiré regions, namely that one in which the Gd atom is on top of a Au atom (denoted TOP), and in the high symmetry hpc/fcc hollow sites of the Au(111) surface (denoted HPC and FCC, respectively).

In Figure 1b, two of the possible adsorption configurations, namely the TOP and FCC, are reported. The latter is almost indistinguishable from HCP one (see Supplementary Information ESI4). These results, obtained by relaxing the coordinates of the outermost three layers, predict that in the region of the superlattice in which the Gd atoms are on top of the substrate Au atoms, the interface structure undergoes significant buckling and deformations involving both the Gd atoms in the alloy and Au(111) surface atoms. In this configuration, the Gd atom displaces inward with respect to the alloy Au atoms by 0.9 Å. Consequently, this pushes the underlying surface Au atoms towards the bulk shifting them by 0.6 Å. Instead, in the hollow configurations, the structural distortions are minor and the planarity of the GdAu$_2$ layer is largely preserved being the Gd atoms higher than the Au atoms of the layer by only 0.2Å (Figure 1b). Further details are given in the supplementary information.

The inward displacement of the Gd and of the underlying Au atoms suggests that the TOP position corresponds to the "dark" areas of the topographic images in analogy to the GdAg$_2$/Ag(111) case[13]. This assignment will be further corroborated by the spectroscopic characterization of the Moiré superstructure described below. As predicted by theory, the two hollow positions are, instead, very similar topographically and spectroscopically and do not allow a clear identification in the topographic image.

The atomic structural relaxation of the TOP regions are reflected in the interfacial binding energy, which our DFT simulations predicts to be larger (i.e. more bound) in the hollow regions than in the top regions by ~11% (-0.087, -0.097, -0.096 eV/Å$^2$ for the TOP, FCC, and HCP, respectively). Topographic images, showing a partial coverage of the GdAu$_2$ on Au(111), support this theoretical prediction (see also Supplementary Information). Indeed, the borders of GdAu$_2$ islands are always formed exclusively by alloy in hollow configuration, i.e. the dark regions of the Moiré superstructure are never observed at domain border, confirming their structural unfavorable energetics. Despite the small energy difference between the TOP and the hollow configurations, extended, continuous alloy layer leading to a Moirè superlattice forms, where TOP, FCC, and HCP regions coexists.

In Figure 2, the local density of states, measured at the TOP and hollow positions of the Moiré superstructure, as indicated by the color-coded circles in Figure 1a, is shown. The spectra present a similar number of electronic features, although local energy shifts and shape variations, observable in the whole energy range, suggest that the density of states is position-dependent. The localization of the density of states at specific surface positions is confirmed by surface energy maps shown in Supplementary Information. A remarkable difference that clearly distinguishes the TOP configuration (black line) from the two hollow geometries (green and red lines) is visible at low energy. Specifically, the TOP regions is characterized by a decrease of the density of states at the Fermi level, which appears depopulated over an energy range of ~0.5 eV (see also inset in panel a). The opening of this gap characterizes the TOP regions, as a finite density of states can be observed in the hollow positions. A peak in the unoccupied energy range is observed at ~700 meV in all Moiré regions although it broadens and splits in two small maxima only in the TOP case. Although comparable in energy, these peaks have different orbital nature. This is suggested by the distinct temperature dependence of the spectra measured in hollow and TOP positions (shown in Supplementary Information ESI-12) and substantiated in the following by theoretical calculations demonstrating the dependence of the electronic structure with the stacking-registry. This highlights that the substrate-induced layer buckling induces a reordering of the orbital character along the Moiré superlattice as well as a band-gap opening in the TOP regions.

In order to provide insight into the measured local electronic structure and to understand its dependence on the crystalline environment, we plot in Figure 3 the calculated total and atom-projected density of states (DOS and PDOS) for the FCC (panel a-c) and TOP (d-f) interface geometries. Furthermore, we calculate the DOS components at the Γ point (Γ-DOS) of the Brillouin zone for the two configurations to reproduce the electron density of states probed by STS. In panel b and e of Figure 3, the gray area marks the total DOS, while the yellow, cyan, and red colors mark the Au, Gd, and Gd-f Γ-PDOS components, respectively.

The most evident difference in these calculated DOS of the TOP and FCC configuration occurs at the Fermi level, where a gap of ~0.5 eV opens in the TOP configuration (Figure 3f), and goes to zero in the FCC configuration (Figure-3 c). This is in agreement with the STS measurements, where a gap is observed only in the "dark" areas of the Moiré superstructure, which are therefore associated to the TOP configuration. Furthermore we observe that in the low energy window [-1.5 eV,1.5 eV], the electron states at Γ have dominant Gd–pd and Au-p orbital components (see S7.1-3). However, a different orbital ordering and character is also evident in the TOP and FCC interfaces for the states around the Fermi level.

We relate the gap opening and these differences in local electronic structure to the out-of-plane displacement of the Gd atoms induced by the adsorption position along the Moirè superlattice. To support this claim we have calculated the electronic structure of free standing $GdAu_2$ layers having the exact distorted structure taken from the TOP and FCC $GdAu_2$/Au(111) calculations. The corresponding Γ-PDOS are displayed in Figure 4. When compared to the Γ-PDOS DOS of a free-standing perfectly-flat $GdAu_2$ layer (see SI-7.1), it is evident that the buckled TOP geometry opens a gap of ~0.25 eV around the Fermi energy, while the gap is not present in the almost flat FCC configuration. This demonstrates that, although the gap size decreases by a factor of two, the gap persists in the distorted TOP geometry even in the absence of the Au(111) substrate. The same analysis also demonstrates that the out-of-plane Gd buckling changes the hybridization of the Gd-pd and Au-sp orbitals having energy close to the Fermi level. In the FCC geometry, the occupied states between -1 eV and the Fermi level are dominated by Gd-pd orbitals while the first two unoccupied states are primarily Au-p states (see SI-7 section). The buckled geometry of the TOP configuration leads to a totally different ordering and character of the electron states between -0.5 eV and 0.5 eV, involving the Au-sp and Gd-pd orbital hybridization (see Figure 4b,d and SI).    This different orbital ordering near the Fermi level, which is clearly controlled by the out-of-plane position of the Gd atom (as confirmed by the free standing calculations), is present also in the surface-supported TOP and FCC GdAu$_2$ layers. We report in the insets above Figure-3 c) and f) the integrated electron density for the states labeled by numbers in the Γ-PDOS plots. In the FCC configuration (Figure-3 c), the Gd–d state 1 is occupied and state 2 is partly occupied. Both of them have clear planar orbital character, which persist up to state 3. The first electron state with substantial out-of-plane orbital component is state 4. Instead, in the TOP configuration, the corresponding Gd-d states are all unoccupied. State 1', the analogue of the occupied state 1 in the FCC configuration, is shifted above the Fermi level at 0.25 eV. In the TOP configuration, the states 2' and 3' - have a marked out-of-plane orbital component. This reordering of the orbital-sequence results from the different valence Gd-pd and Au-sp hybridization that occur when the Gd atoms displace out of the layer plane.

Quite interestingly, the layer buckling controls also the energy of the occupied and unoccupied Gd-f states (see PDOS of Figures 3 and Table SI-5.2 in the SI). In the supported GdAu$_2$ layer, the occupied Gd-f band center of mass lowers by ~0.25 eV going from the FCC to the TOP geometry. This results in a dependence of the energy of the f multiplet according to the adsorption configurations as highlighted in Figure 5. Here, we plot the energy $E_f$ of the Gd-f states, calculated as a weighted average from the PDOS, as a function of the structural parameter δz, i.e. the out-of-plane position of the Gd atoms. Values of |δz| equal to 0, 0.2Å and 0.9Å correspond to the geometry of a perfectly flat, and of the buckled FCC and TOP GdAu$_2$ layers. To further corroborate our claim, for the free-standing case we calculated an additional geometry displacing the Gd atom by |δz|=1.5 Å. The data for the supported layer (black symbols and lines) demonstrate that the energy shift $E_f$ is proportional to |δz| and therefore that it is controlled by the layer buckling. Noteworthy, the data for the corresponding free-standing distorted configurations (red symbols and lines) follow the same trend. Moreover, this also demonstrates that the relative energy difference of the Gd-f states between TOP and FCC configurations is induced by the layer buckling while the coupling to the supporting Au crystal induces an almost rigid shift of 0.6-0.8eV of the Gd-f states in both the TOP and FCC configurations.

The spatial modulation of the f states can indeed be clearly observed in the experimental measurements. In Figure 6, we compare the density of states of the monolayer GdAu$_2$ on Au(111) measured with space averaging techniques, as Ultra-Violet (UPS) and Inverse Photoemission Spectroscopy (IPES), with local spectroscopic measurements. The most distinguishing feature that characterizes the occupied electron states of the supported GdAu$_2$ from the Au(111) electronic density of states (panel a, blue and yellow lines, respectively) is a peak visible at about 10 eV below the Fermi level and corresponding to the Gd 4f electron state[14-15]. The features of the IPES spectrum measured on the GdAu$_2$/Au(111) interface are clearly distinguishable from the ones measured on the clean Au(111) substrate (panel b) and closely resemble the one observed in the dI/dV curves (Figure 2). In figure 6b, the arrow highlights the highest energy feature of the IPES spectrum, which corresponds to the photon emission from the unoccupied Gd 4f multiplet states (See also Sup.Info). Summarizing these UPS-IPES spectra we noticed that both occupied and unoccupied Gd 4f states are observed at energies comparable, but slightly shifted, to the ones reported in literature for Gd and GdAu$_2$ bulk material. This feature is, however, considerably broadened[14, 16-19]. In a previous work, we have reported that the formation of the GdAu$_2$ alloy does not change the occupancy of the 4f multiplet with respect to the Gd bulk, which remains half occupied[12]. This confirms the expected low reactivity of the 4f electrons, which do not participate directly to chemical bonds and are shielded by the more delocalized s and d electrons. Despite this, the structural changes and re-hybridisation of the d orbitals previously described can cause small perturbations to the 4f multiplet even if their occupation number is unaffected.

In order to clarify this we have probed the energy region of the 4f unoccupied electrons with scanning tunneling spectroscopy (STS). This region is energetically accessible, on difference to the occupied 4f states, however, the highly localized nature and the small radial expansion of the 4f electrons result in a rather weak dI/dV signal [20]. The spectra recorded at the various positions of the Moiré superstructure are reported in Figure 6c. For the sake of completeness, in panel c, we report also the STS spectrum measured on the Au(111) surface. The small feature measured on this surface coincides with the reported onset of the Au bulk empty states[21], which is not visible on the alloy overlayer. The onset of the f states, indicated by arrows, is relatively shifted of about 0.2-0.3eV from TOP to hollow position, being the first the lowest in energy. This result, which is in full agreement with the predictions shown in Figures 3 and 5, suggests that energy-shifted contributions sum in the UPS-IPES spectra leading to its broadening.

The shift of the 4f states of Gd, should not be surprising. Indeed, although these states do not participate directly to chemical bond, they are sensitive to the chemical environment through the screening effect of the valence band electrons as observed in pure Gd in PES and IPES spectra[14, 16-18, 22]. In DFT+U calculations, the chemical composition and crystalline environment is accounted by the parameter U, whose value influences the computed energy of the f states. In our calculations, in order to disentangle the effect of the layer buckling on the Gd-f energy shift, we followed previous works and set the value of $U_{eff}$ for the GdAu$_2$ alloy equal to that suggested for Gd bulk[23-24]. The value of $U_{eff}$ reflects the on-site Coulomb interaction of the localized f electrons, which is likely to be different in the GdAu$_2$ alloy and in Gd bulk. As a result, comparing the absolute position of the Gd-f states in the bulk and in the alloy is not meaningful in within our DFT+U approach. The relative Gd-f energy shifts arising from the out-of-plane GdAu$_2$ distortions should instead be well captured by the calculations. As shown in the Supplementary information (ESI-6), using larger values of this on-site parameter ($U_{eff}$~7eV), corresponding to higher Gd-f electron Coulomb repulsion in the GdAu$_2$ layer than in Gd bulk, would shift the energy of the Gd-f states to lower energies, in better agreement with the measured absolute energies of the occupied states ~-10 eV (see below).

We can discard the contribution of charge-transfer effects from the observed energy shift according to the results of the Bader charge analysis (see Supplementary Information ESI-8). This shows that, with respect to isolated atoms, the charge rearrangement occurs almost exclusively between the Gd atom and the Au atoms of the alloy layer, reducing the 28$e$ in the elemental bulk Gd, to 26.9$e$ and to 26.8$e$ in the free standing and supported GdAu$_2$ monolayer, respectively. The supported GdAu$_2$ layer is slightly positive but its overall charge is basically preserved with respect to the free-standing value while the alloy layer has a partial ionic character. Therefore, we conclude that the physical origin of the predicted and experimentally observed energy shift of the 4f-states across the Moiré lattice is not induced by charge transfer effects but by the structural layer buckling experienced by the Gd atoms in the different adsorption positions.

The present work provides valuable insights into the correlation between the electronic properties of 2D layers and its planarity. The stacking configuration and the structural relaxation prompt, despite their relatively weak interaction, a considerable local atomic buckling in the GdAu$_2$ superstructure on Au(111) surface and consequently a clear patterning of the electronic density of states with nanoscale spatial periodicity.

Given the described modulation of the electronic properties, it can be expected that this layer acts as a chemical template. To test this hypothesis, we have performed a set of preliminary calculations by adsorbing a probe H atom on the top of Au and Gd atoms of the TOP and FCC configurations of the Au(111)-supported GdAu$_2$ system. Indeed, these calculations demonstrate the patterning of the surface reactivity, with a preferential binding of H to the Au atoms of the FCC superlattice regions with respect to the Au atoms of the TOP regions. Despite the high H coverage used in these model simulations, the difference in the calculated adsorption energies for the different supercell regions is larger than 0.3 eV/atom. These findings therefore provide valuable insights into the correlation between the electronic properties of 2D layers and its planarity and a perspective for controlling the electronic, magnetic and reactivity properties in atomically-thin layers.

## Acknowledgements

This project is partially funded by the EU-H2020 research and innovation programme under grant agreement No 654360 NFFA-Europe and the Spanish ministry of Economy (MAT2013-46593-C6-4-P and MAT2016-78293-C6-5-R)
We are grateful to Prof. S. Modesti (CNR-IOM) for his support with the electronics. Federico Salvador and Andrea Martin, are kindly acknowledged for technical assistance.

## Notes and references

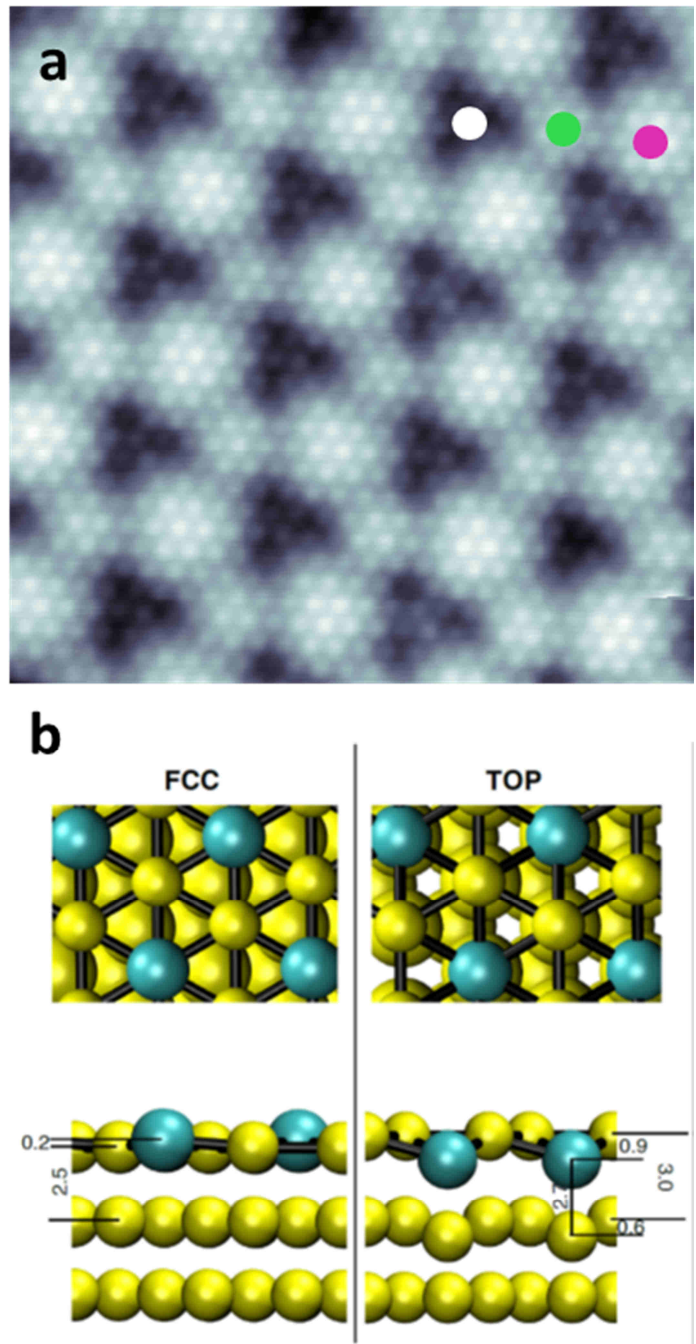

**Figure 1: Structural characterization of GdAu$_2$/Au(111)** a) Topographic image of the GdAu$_2$/Au(111) (12nmx12nm). b) Calculated equilibrium geometries of the GdAu$_2$/Au(111) interfaces in the FCC and TOP regions. Gd and Au atoms are represented by cyan and yellow colors. Only the two outermost Au(111) layers are displayed. The numbers report selected vertical distances (in Å) that characterize the interface distortions. The dots indicated in panel a) refer to 3 different regions TOP, hollow FCC and HCP stacking positions of the Moiré superstructure (see text).

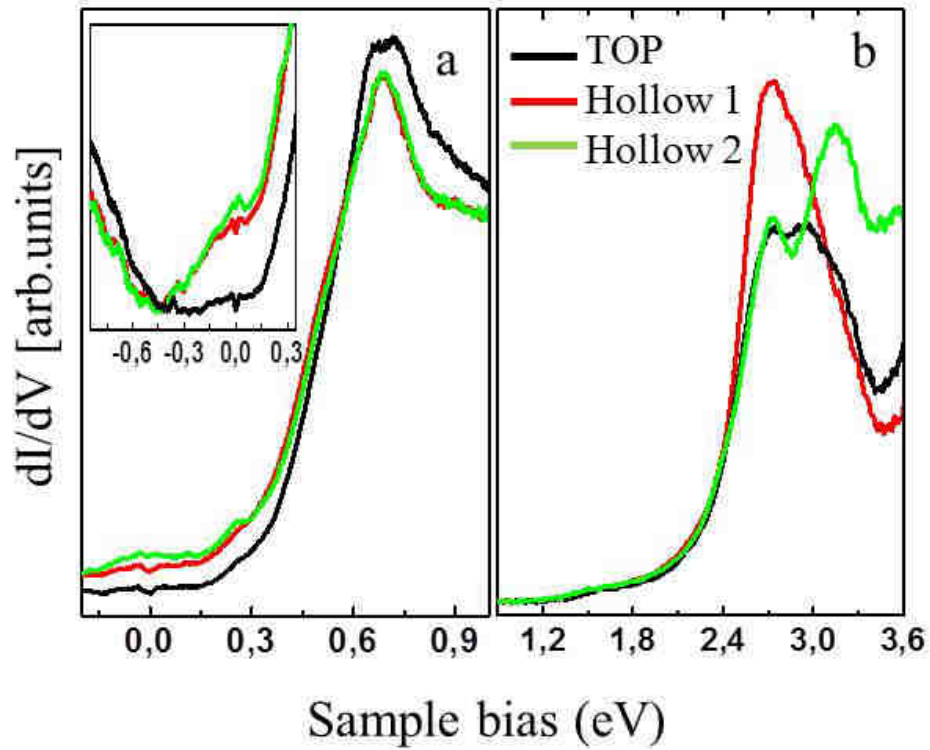

**Figure 2: Local spectroscopy at different positions on the Moiré superstructure**. The spectra have been measured at the 3 characteristic positions of the Moiré pattern according to the colour coded circles indicated in Figure 1a. These are shown in separated windows to highlights the low energy contributions, which have a small intensity. The inset in panel a shows the density of states at the Fermi energy.

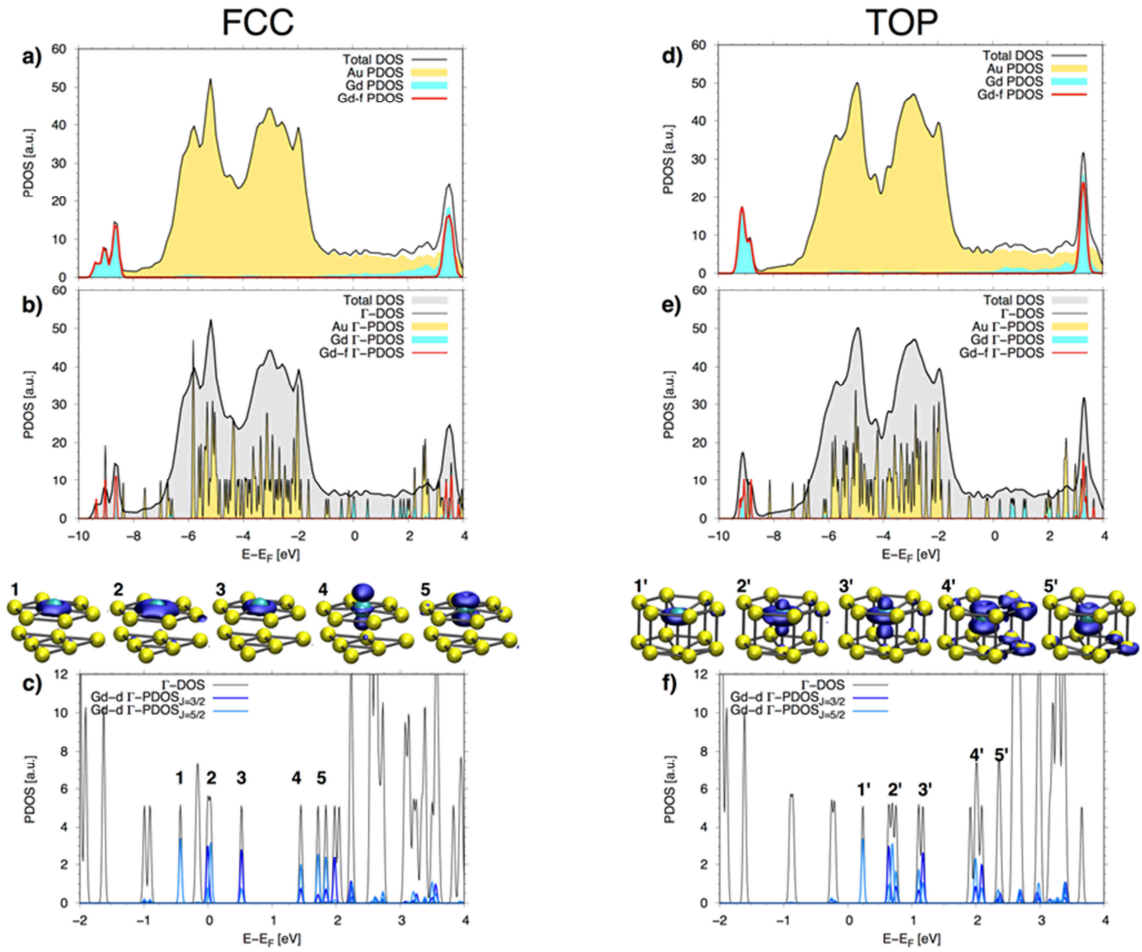

**Figure 3: (2 columns) DFT electronic structure of GdAu$_2$/Au(111) in FCC (left) and TOP (right) configurations.** a,d) Total and atom-projected density of states (DOS and PDOS). b,e) DOS components at the Γ point of the Brillouin zone (Γ-DOS). c,f) Orbital analysis and charge density (insets) corresponding to the states labelled by numbers.

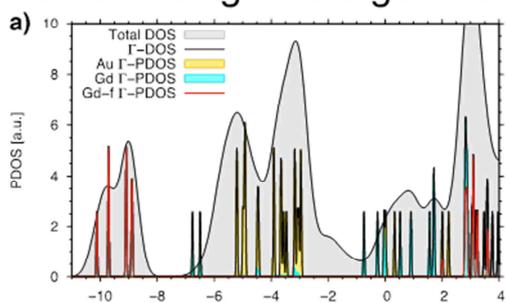
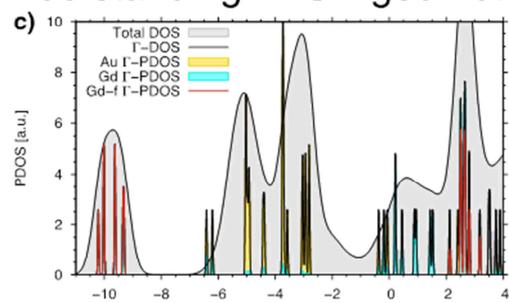
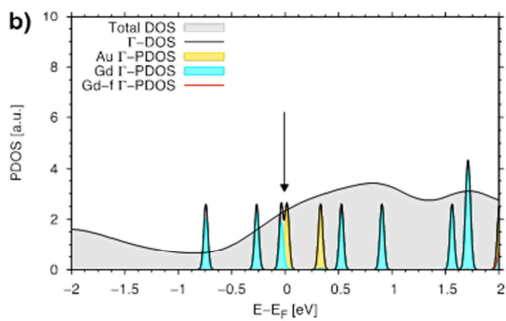
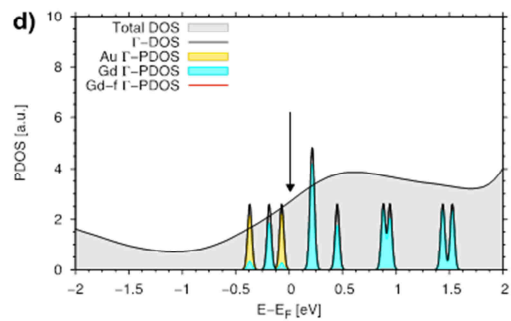

**Figure 4: Effects of buckling on the electronic structure.** Calculated DOS, and Γ-PDOS for free-standing GdAu$_2$ layers with structural distortions as in the FCC (left) and TOP (right) supported configurations.

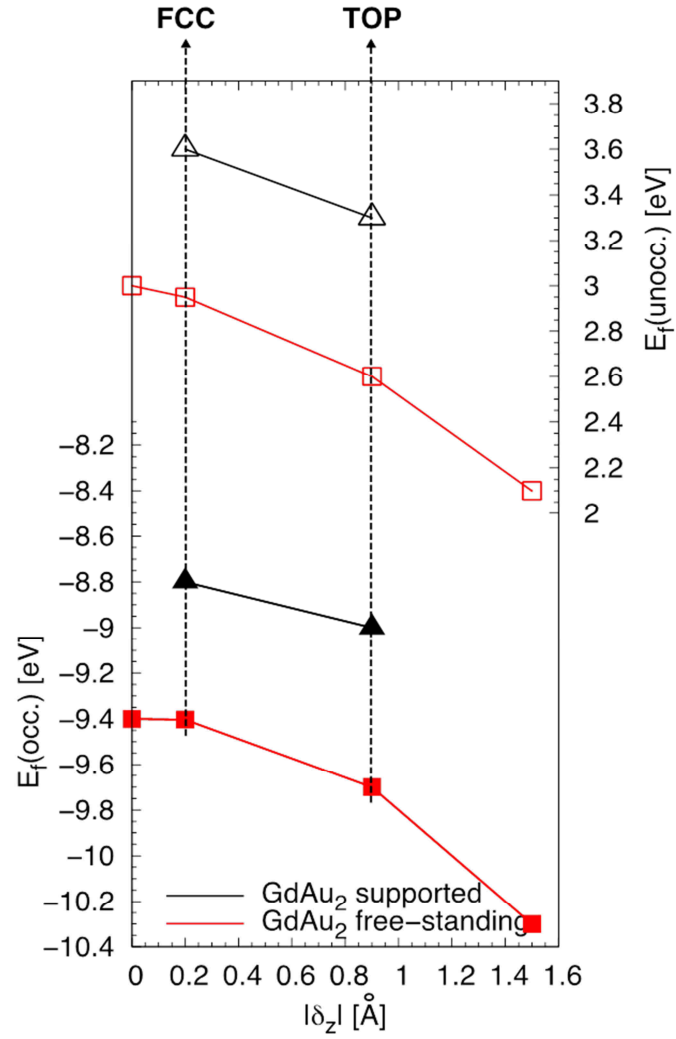

**Figure 5. Induced energy shift of the center of mass of the Gd 4f multiplet states as a function of the Gd atom inward displacement in GdAu$_2$ layers.** Filled and empty symbols represent occupied states (left energy scale) and unoccupied states (right energy scale).

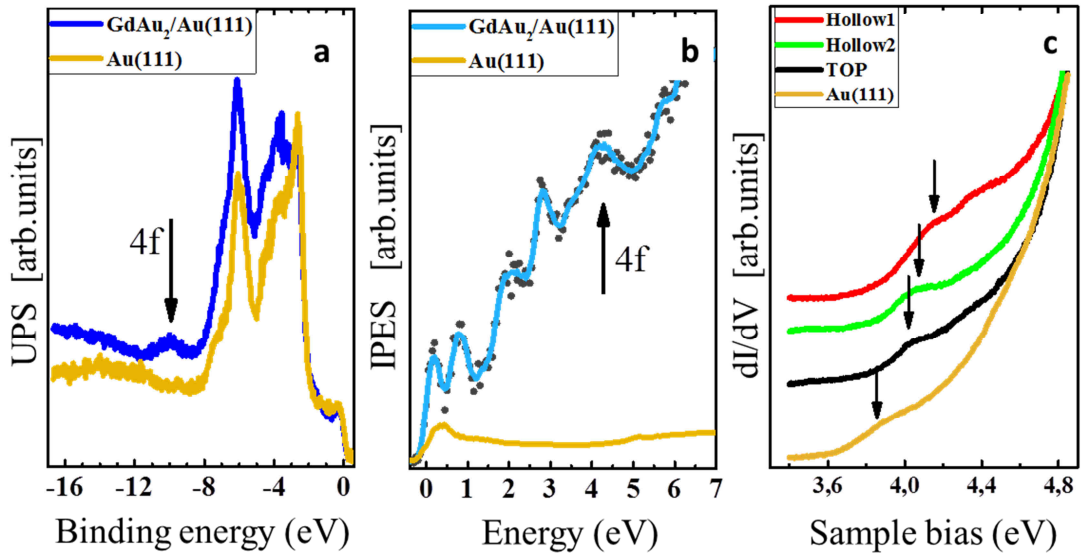

**Figure 6. Spectroscopic measurements of the Gd 4f multiplet on the GdAu$_2$/Au(111) interface.** a) Occupied density of states (HeII), coverage 0.6ML; b) unoccupied density of states (IPES), coverage 0.9ML; c) Local spectroscopic measurements of the unoccupied 4f density of states at the indicated position on the Moiré superstructure. The spectra are displaced for better visualization. The arrows indicate the onset of the 4f multiplet and of the onset of the Au(111) bulk band (ref.21), respectively.

# Electronic Supplementary Information

# Self- texturizing electronic-properties of a 2-dimensional GdAu$_2$ layer on Au(111): the role of out-of-plane atomic displacement

*Alexander Correa[1,2,3], Matteo Farnesi Camellone[4], Ana Barragan[1,3], Abhishek Kumar[5,6], Cinzia Cepek[5], Maddalena Pedio[5], Stefano Fabris[4], Lucia Vitali[1,3,7]*

[1] Departamento de física de materiales, Universidad del País Vasco, ES-20018 San Sebastián (Spain)

[2] Donostia International Physics Center, 20018 San Sebastián (Spain)

[3] Centro de Fisica de Materiales (CSIC-UPV/EHU) y Material Physics Center, ES-20018 San Sebastián (Spain)

[4] CNR-IOM DEMOCRITOS, Istituto Officina dei Materiali, Consiglio Nazionale delle Ricerche and SISSA, Via Bonomea 265, I-34136, Trieste, Italy

[5] Istituto Officina Materiali (CNR-IOM), Laboratorio TASC, I-34149 Trieste (Italy)

[6] Dipartimento di Fisica, Università di Trieste, , I-34127 Trieste (Italy)

[7] Ikerbasque Foundation for Science, ES-48013 Bilbao (Spain)

1. **Sample preparation**

   The GdAu$_2$ surface alloy is achieved depositing in ultra-high vacuum conditions (UHV) via electron bombardment gadolinium atoms onto the Au(111) surface previously prepared by cycles of Ar+ ion sputtering and subsequent annealing. The Au(111) surface held at a temperature between 280º and 360ºC ensures the formation of ordered GdAu$_2$ alloy layers. For the realization of this work, the surface was prepared on three different UHV systems. The photoemission and inverse photoemission experiments were performed at room temperature, while the scanning tunneling spectroscopy was performed on the sample hold at 1K in a bath cryostat.

2. **Intermetallic structure**

   The GdAu$_2$ alloy structure was identified as by Corso et al. [1]. According to this work the unit cell of the alloy layer is described in the inset of figure ESI2.1, where the blue circles and the yellow ones represent the Gd and Au atoms of the unit cell respectively.

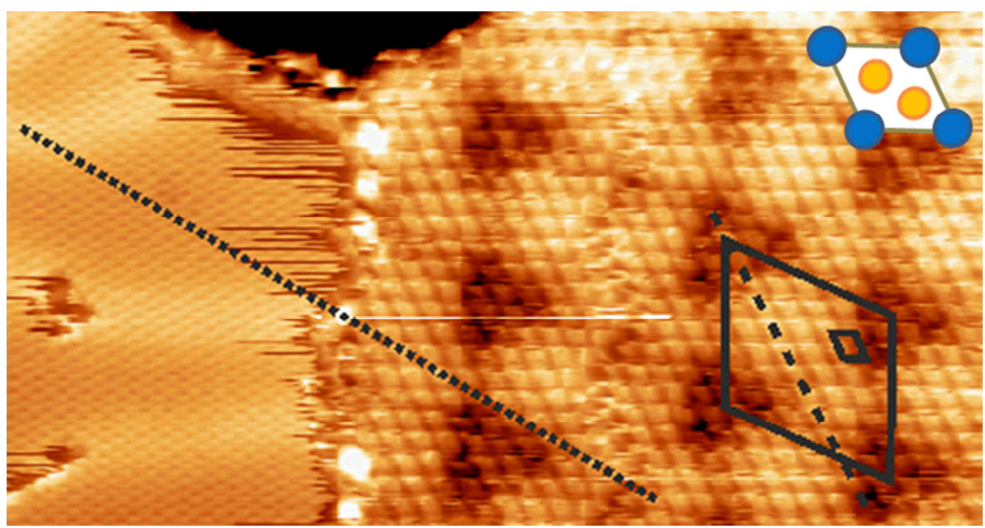

**Figure ESI2.1. Structure of GdAu$_2$/Au(111).** Topographic image of a partial coverage of GdAu$_2$ on Au(111) showing atomic resolution. Image size 20nm x 11nm. The dashed and dotted lines indicate high symmetry directions of the moiré and Au(111), respectively. The small and large rhombi indicate the alloy and superstructure unit cells, respectively. In the inset, the proposed structure of the GdAu$_2$ alloy.

Scanning tunneling microscopy images showing simultaneous atomic resolution in the Au(111) surface and in the alloy layer allow determining the angle of rotation of the moiré superstructure and of the alloy layer with respect to the Au(111) surface, as 33º and 5º respectively. Similarly, the periodicity of the alloy and of the moiré superstructure unit cell, depicted with rhombi in figure ESI2.1, have been determined to be 5.3Å and of 37Å respectively. Minor local variations of the supercell size can be detected. These do not affect the density of states as described for the similar structure $GdAg_2$/Ag(111) [2] suggesting a different interaction with the substrate.

It is also worth noticing that the border of alloy islands are always terminated by Gd in hollow stacking position with respect to the underlying Au(111) (see main text). Indeed, the apparent topographic minima, i.e. the darker areas of the moiré pattern, are energetically more expensive as described in the main text.

3. **Computational methods and supercells structure**

All the DFT simulations were carried out at the GGA[3] level using the plane-wave Quantum-Espresso[4] code. The spin-polarized Kohn-Sham equations were solved within the plane wave/pseudopotential framework using Vanderbilt's ultrasoft pseudopotentials[5], employing basis-set cutoffs of 117 Ry and 579 Ry for the electron wave function and density, respectively. The Brillouin zone of all $GdAu_2$/Au(111) computational supercells was sampled with a regular mesh of 8x8 k points. A Hubbard U [6] term acting on the Gd-4f orbitals was added to the GGA energy functional (GGA+U) in all calculations. The values of the U and J parameters (U=6.7 eV J=0.7 eV) were chosen following previous works, so as to provide a reasonable representation of the f-bands energy position in Gd bulk [7-8]. Structural relaxations of the interfaces were performed at the GGA+U level. Here, only the atoms of the three bottom Au layers were kept fixed, while all other atoms were free to move during optimization. The electronic properties were calculated starting from these optimized geometries and performing an additional set of GGA+U calculations that included the spin-orbit interaction. These last calculations were used for the analysis of the electronic structure, the total and projected density of states, as well as for the interface energetics.

The GdAu$_2$/Au(111) system was modeled with a hexagonal supercell having a (√3x√3)R30º orientation with respect to the (1x1) surface lattice of the Au(111) substrate (Fig. ESI3.1). The in-plane lattice constant of the supercell was 5.11 Å, while the out-of-plane one allowed for a region of vacuum between periodic images larger than 10 Å.

The Au(111) substrate was modeled with 5 atomic layers. We considered three different stacking configurations of the GdAu$_2$ layer with respect to Au(111) substrate. The three stacking configurations are defined as TOP, HPC and FCC following the position of the Gd atoms with respect to the high-symmetry sites of the underlying Au(111) surface.

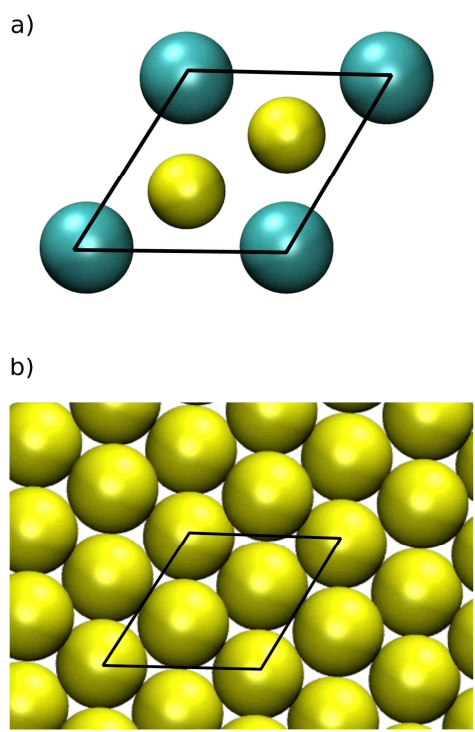

**Figure ESI3.1 Computational cells:** the GdAu$_2$ layer (a) and the supporting Au(111) surface (b).

## 4. FCC and HCP interface structures

Here we provide evidence that the optimized geometries of the HCP and FCC interface configurations are almost identical. The corresponding structures are shown in Figure ESI4.1, which displays the side and top views of the outermost three layers. The differences in the interfacial vertical distances between the FCC and HCP configurations are lower than 0.01 Å, i.e. smaller than the error bars in DFT calculations (Table ESI4.2).

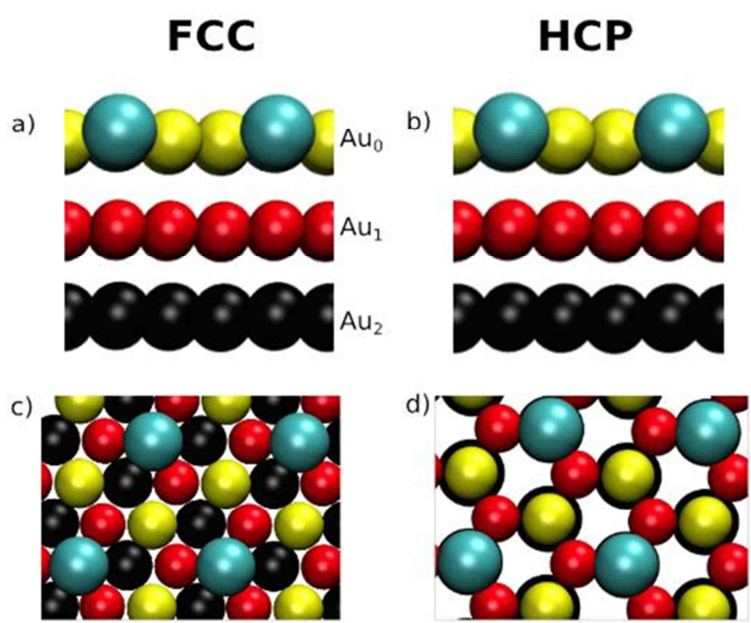

**Figure ESI4.1 Optimized structural geometry.** Upper panel: Side view of the FCC (a) and HCP (b) geometries. Lower panel: Top view of the FCC (c) and HCP (d) geometries. Only the first three layers are shown. Yellow, red and black colors denotes the Au atoms in the GdAu$_2$ alloy, and in the first and second Au(111) layers, respectively.

|  | TOP | FCC | HCP |
|---|---|---|---|
| $d_z$(Gd-Au$_1$) | 2.8 Å* | 2.7 Å | 2.7 Å |
| $d_z$(Au$_0$-Au$_1$) | 3.0 Å | 2.5 Å | 2.5 Å |
| $d_z$(Au$_1$-Au$_2$) | 2.4 Å | 2.4 Å | 2.4 Å |

**Table ESI4.2. Vertical distances between Gd and Au atoms in the 3 interface configurations.** Au atoms are labeled according to their layer position, see Figure ESI4.1. * In the TOP configuration dz represents the bondlength between the Gd atom and the underlying atom of the Au1 layer, which is highly distorted and non-planar.

## 5. Energy of the Gd-f states

To understand the dependence of the Gd f states on the atomic environment, we have computed the Gd-f projected DOS for the following systems: Gd bulk, free standing GdAu$_2$, and GdAu$_2$/Au(111) interfaces in TOP, FCC and HCP configurations. These PDOS are shown in Fig. S5.1. The Gd-f PDOS was used to compute the weighted mean energy of the Gd f states in these different systems. The values are reported in Table ESI5.2.

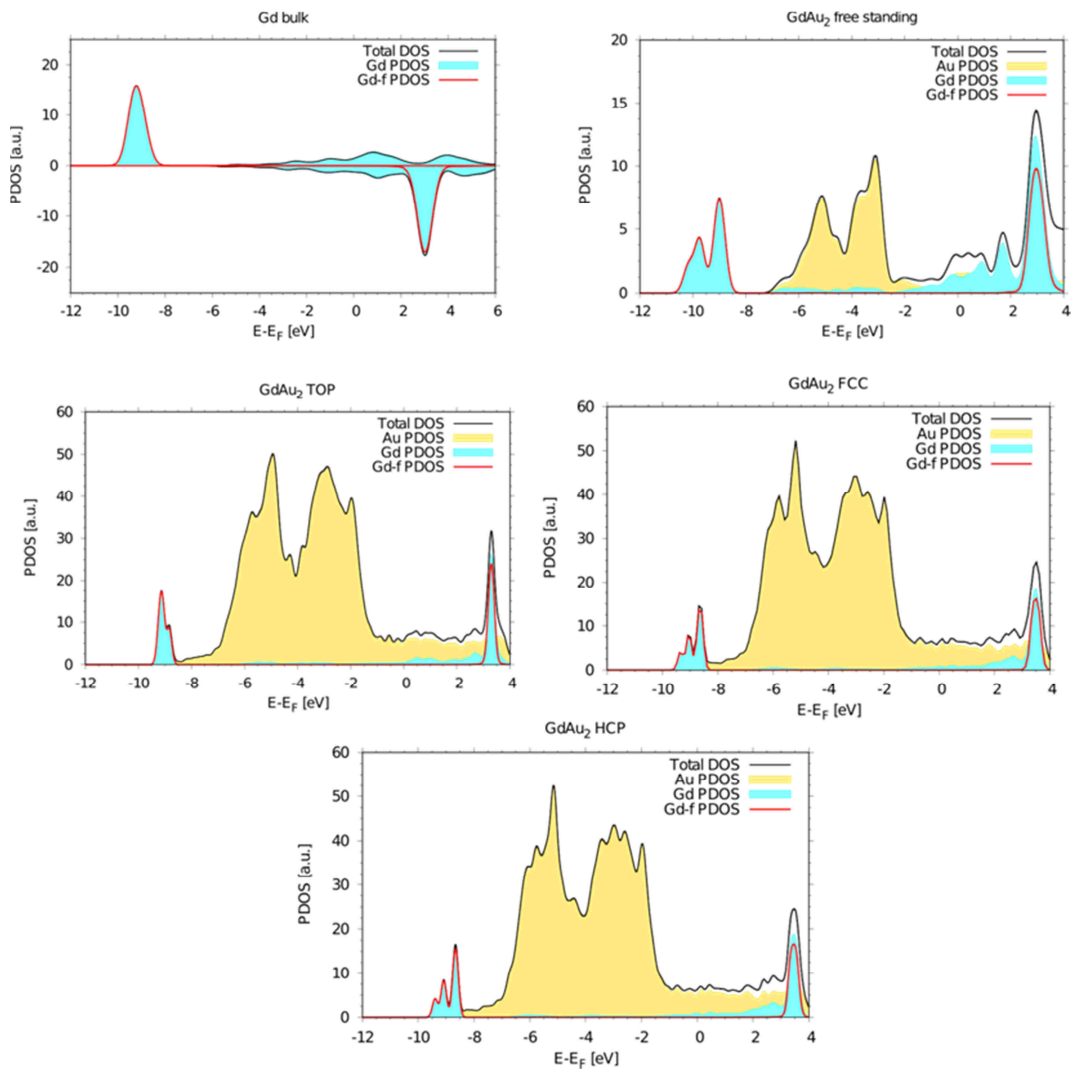

**Figure ESI5.1. Computed total and atom-Projected density of states (DOS and PDOS).** Various systems as bulk Gd, free standing GdAu$_2$ layer, TOP, FCC and HCP GdAu$_2$/Au(111), are compared.

| Gd-f states | Gd Bulk | GdAu$_2$ free-standing | TOP | FCC | HCP |
|---|---|---|---|---|---|
| Occupied | -9.2 eV | -9.4 eV | -9.0 eV | -8.8 eV | -8.9 eV |
| Unoccupied | 3.0 eV | 3.0 eV | 3.3 eV | 3.6 eV | 3.5 eV |

**Table ESI5.2. Mean Energy of the occupied and unoccupied Gd f-states.** The values of Gd bulk, free standing GdAu$_2$, and GdAu$_2$/Au(111) interfaces in TOP, FCC and HCP configurations are given.

## 6. U dependence

In within the DFT+U approach, the energy of the Gd-f states, on top of being determined by the chemical and crystalline environments, is also controlled by the value of the effective parameter $U_{eff}$=U-J. We have performed additional calculations to check how this parameter affects the electronic structure of the GdAu$_2$/Au(111) system and in particular the dependence of the Gd-f energies. This analysis was performed on the TOP configuration only, since the same effect is expected for the other ones. We have considered three values of $U_{eff}$: $U_{eff}$=5, 6, 7 eV. The resulting DOS and PDOS analysis are plotted in Fig ESI6.1. The position of the weighed mean energies of the occupied and unoccupied 4f states are collected in Table S6.2. This analysis shows that a better agreement with the experimental position of the f states in GdAu$_2$ is obtained by using values of $U_{eff}$~7 eV, therefore larger than those reported for Gd bulk.

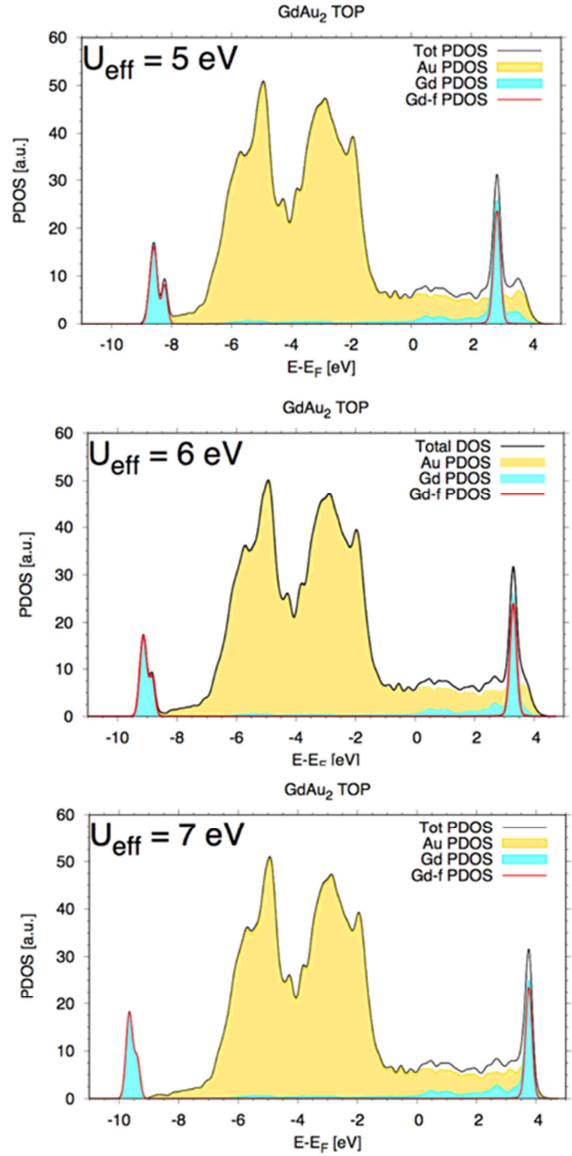

**Figure ESI6.1 Dependence of the energy of the Gd f states on the value of $U_{eff}$.** The $GdAu_2/Au(111)/Au(111)$ TOP interface is taken as a representative case.

| Gd-f states | TOP $U_{eff}$ = 5 eV | TOP $U_{eff}$ = 6 eV | TOP $U_{eff}$ = 7 eV |
|---|---|---|---|
| Unoccupied | 2.8 eV | 3.3 eV | 3.7 eV |
| Occupied | -8.5 eV | -9.0 eV | -9.6 eV |

**Table ESI6.2. Weighed mean energy values of the occupied/unoccupied 4f states**. The mean values of $GdAu_2/Au(111)/Au(111)$ TOP interface as a function of U is given.

## 7. Orbital Analysis

A detailed orbital analysis in terms of the Γ–PDOS for the s, p, and d orbitals of the Gd and Au atoms is presented in Figure ESI7.1 for the free-standing flat $GdAu_2$ layer, in Figure ESI7.3 for the supported $GdAu_2$/Au(111) system in TOP and FCC configurations (the PDOS refer to the outermost layer only), and in Figure ESI7.2 for free-standing $GdAu_2$ layers having the same structural distortions of the supported TOP and FCC configurations. These data shows that 1) the $GdAu_2$ layer electron states laying in the [-2 eV,2eV] region across the Fermi level have primarily Au-p and Gd-d character (Figures ESI7.1-3); 2) in the free-standing layers, with respect to the flat geometry, the TOP and FCC structural distortions induce shifts and reordering of the electron states (Fig. S7.2) and 3) the coupling with the substrate increases the width of all the bands having Au-d, Au-p, Gd-d and Gd-p character (Figures ESI7.3).

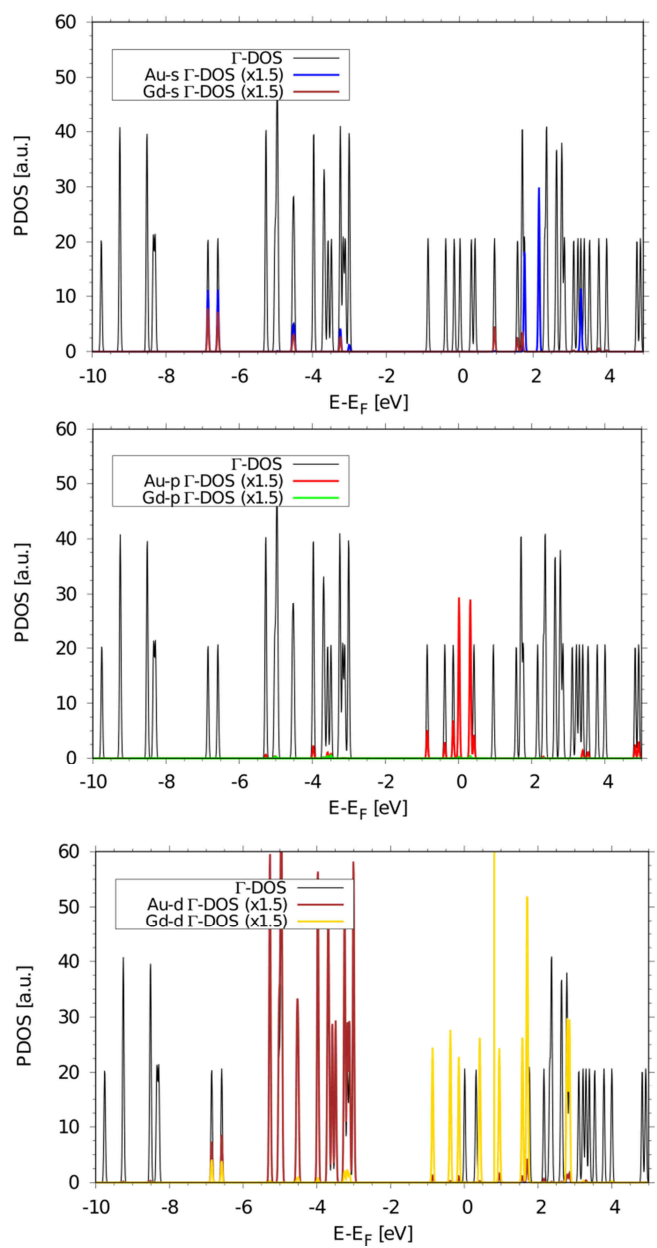

**Figure ESI7.1 Total and Projected density of states at Gamma (Γ–DOS and Γ–PDOS).** The s, p, and d orbitals of the Gd and Au atoms of a free-standing $GdAu_2$ flat layer are shown.

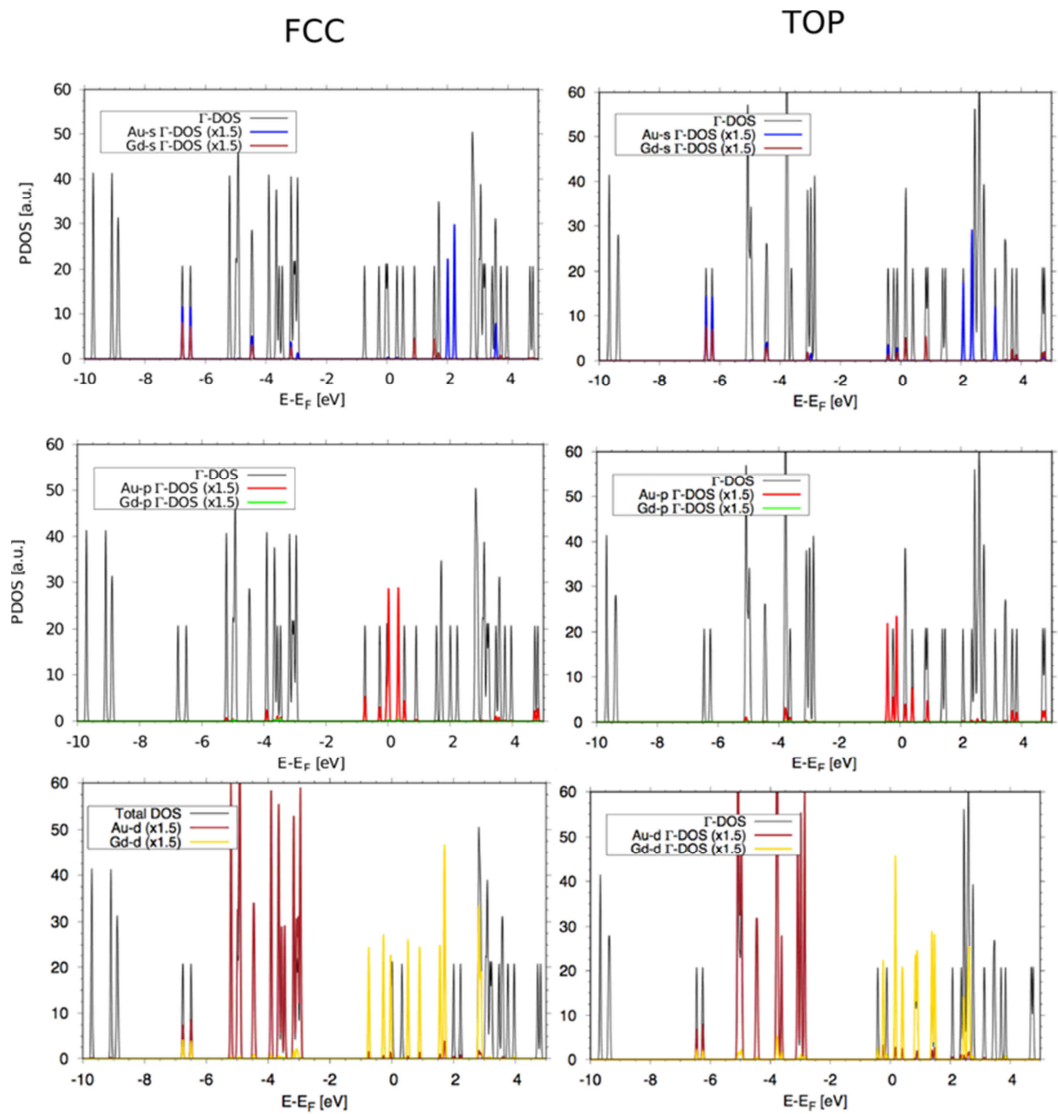

**Figure ESI7.2. Γ–DOS and Γ–PDOS for the s, p, and d orbitals of the Gd and Au atoms of free-standing GdAu$_2$ layer.** The structural distortions of the atoms in the free standing layer are the same as in the supported FCC and TOP configurations.

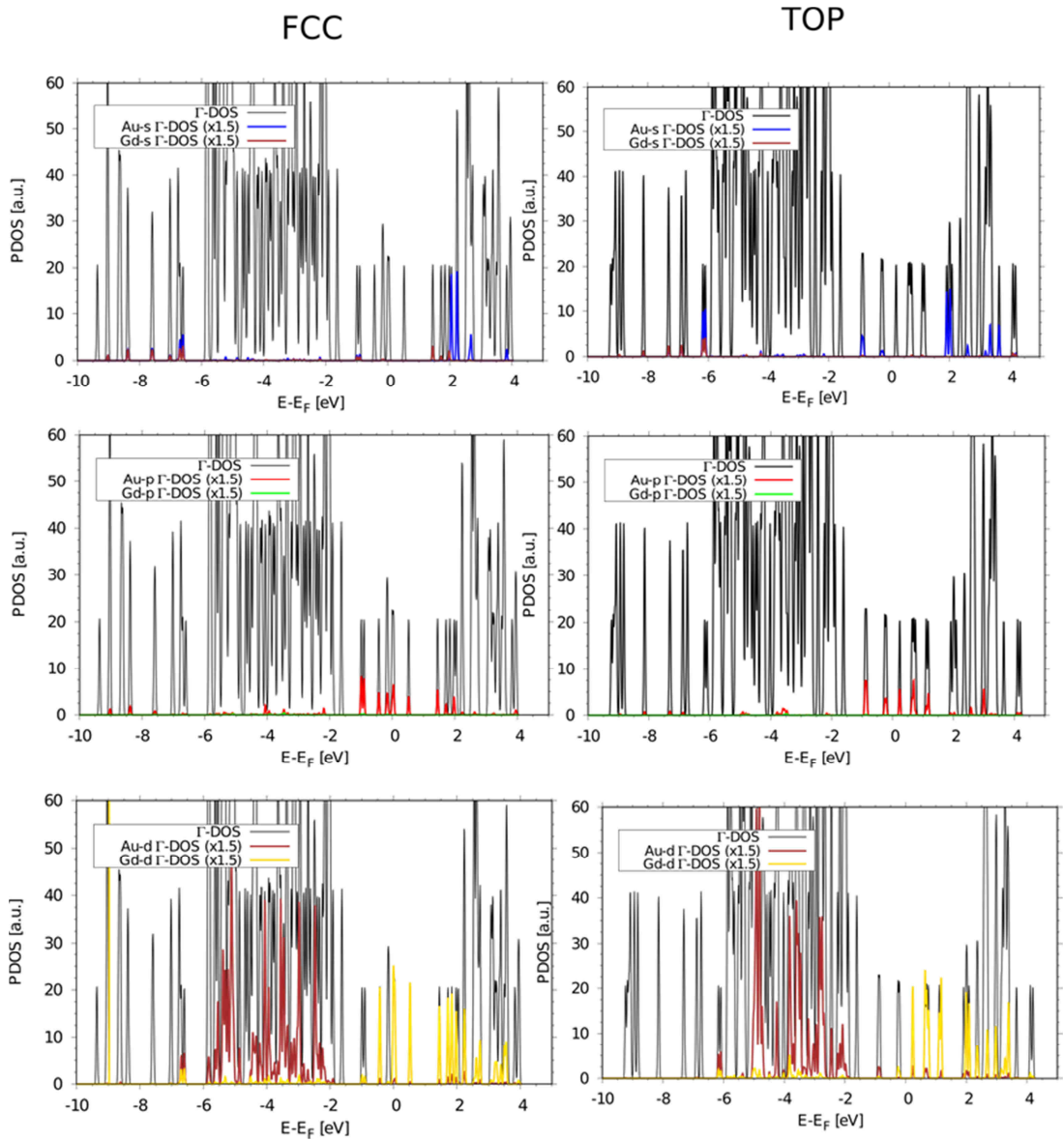

**Figure ESI7.3. Γ–DOS and Γ–PDOS of supported GdAu$_2$/Au(111) system:** the s, p, and d orbitals of the Gd and Au atoms in the FCC and TOP configurations are shown in separated panels.

## 8. Bader charge analysis

The Bader charge analysis for a set of different systems is presented in Table ESI8.1. This data demonstrates the partial ionic character of the Gd-Au bond in the GdAu$_2$ alloy and that the charge transfer at the GdAu$_2$/Au(111) interface is negligible.

|    | Au Bulk | Gd bulk | GdAu$_2$ Free standing | GdAu$_2$/Au TOP | GdAu$_2$/Au FCC | GdAu$_2$/Au HCP |
|----|---------|---------|------------------------|-----------------|-----------------|-----------------|
| Gd |         | 28 $e$  | 26.89 $e$              | 26.84 $e$       | 26.87 $e$       | 26.88 $e$       |
| Au | 11 $e$  |         | 11.55 $e$              | 11.39 $e$       | 11.43 $e$       | 11.43 $e$       |

**Table ESI8.1. Bader charges computed for the Gd and Au bulk system**. Free-standing GdAu$_2$ layer, and for the GdAu$_2$/Au(111) TOP, FCC and HCP interfaces.

## 9. Ultra violet photoemission spectroscopy method

Ultraviolet VB photoemission spectra were acquired in an ultra high vacuum (UHV) experimental apparatus (base pressure ~4x10$^{-10}$ mbar), using the He II emission of a conventional He discharge lamp ($h\nu$ = 40.8 eV) and a hemispherical electron energy analyzer (PSP). Electron-energy distribution curves were measured at room temperature in normal emission geometry, with an overall instrumental energy resolution of ~0.25eV. All binding energy values are referred to the Fermi energy level of the clean gold crystal.

## 10. Inverse Photoemission electron spectroscopy

The Normal Incidence IPS measurements were performed in the UHV system of IOM SIPE laboratory by using an Erdman-Zipf electron gun, with the electron beam divergence < 3°. Photons emitted from the sample surface are collected by a Geiger-Müller detector with a He-I$_2$ gas mixture and a SrF$_2$ entrance window filtering photons at energy hv = 9.5 eV. Current on the sample was < 1 µA. The overall resolution was < 300 meV, as measured by the Fermi level onset of a clean Ta foil.

Spectra are normalized at each point to the incident electron beam current. Measurements have been obtained at normal incidence, at room temperature.

The 4f unoccupied states show strong cross section dependence from the detected photon energy [9]. Nevertheless the filled Gd 4f states present a binding energy close to the detection photon energy of our Geiger Mueller detector (9.5 eV). The increase of the cross section is due to the resonance effect which allows our assignment[10]. The observed Gd 4f empty states at about 4 eV are fully compatible with the values reported in literature [11].

11. **Map of density of state**

In Figure ESI11.1, topographic and density of states maps achieved simultaneously at selected energies are shown. Aim of this is the identification of the spatial localization of the density of states maxima. This allows distinguishing the major contribution of TOP and HOLLOW sites in the total density of states as highlighted by the corrugation profile (right column).

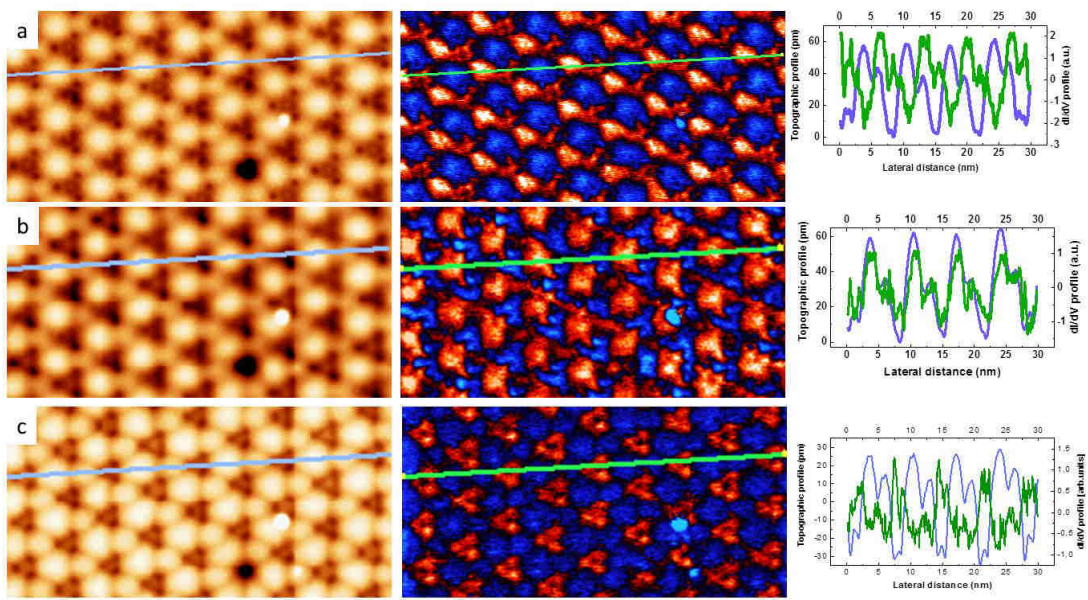

**Figure ESI11.1: Comparison of topographic and density of states maps at defined energy.** Topographic images (left columns) and corresponding density of sates maps at the indicated energy (a) 3.2eV; (b) 2,75eV; (c) 650meV. Image size: 30nm x 15nm. The blue and green lines highlight the spatial correlation between the topographic positions and the density of state maxima, respectively.

12. **Temperature effects on the electron density of states**

In Figure ESI12.1, we show the temperature evolution of the low energy density of states at TOP and HOLLOW position in the GdAu$_2$/Au(111). It is worth noting that the peak which appears split in TOP position at low temperature becomes a sharp peak as the temperature increases (bottom to top curve). Vice versa, the feature observe in the HOLLOW positions broadens and split as a function of increasing temperature. This suggests that although the feature is observed at the same energy, the orbital character of this feature differs in the two cases highlighting the role of the crystal field in the two atomic stacking configurations.

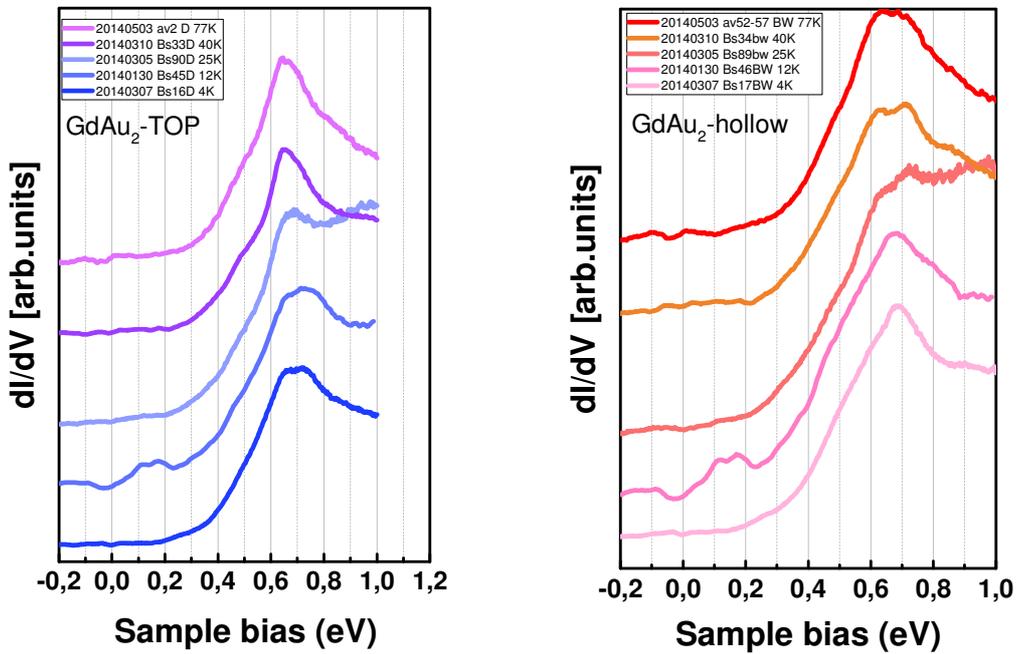

**Figure ESI12.1. Temperature of effect on the density of states.** Series of spectra showing the different behavior of the low energy peak of GdAu$_2$/Au(111) as a function of increasing temperature (from bottom to top).